\documentclass[prl,superscriptaddress,twocolumn,showpacs,%
preprintnumbers,amsmath,amssymb]{revtex4-1}
\usepackage{graphicx}
\usepackage[breaklinks, colorlinks=true, pdfstartview=FitV,%
 linkcolor=red, citecolor=blue, urlcolor=blue]{hyperref}
\usepackage{color}

\newcommand{\tr}{\text{tr}}
\newcommand{\rme}{\text{e}}
\newcommand{\rmi}{\text{i}}
\newcommand{\Nc}{N_{\text{c}}}

\newcommand{\muq}{\mu_{\text{q}}}
\newcommand{\muB}{\mu_{\text{B}}}
\newcommand{\tmu}{\tilde{\mu}_{\text{q}}}
\newcommand{\calM}{\mathcal{M}}
\newcommand{\calN}{\mathcal{N}}
\newcommand{\calL}{\mathcal{L}}
\newcommand{\calP}{\mathcal{P}}

\newcommand{\Pol}{L}

\begin{document}
\preprint{RIKEN-QHP-94}
\title{Sign problem and the chiral spiral on the finite-density lattice}

\author{Ryutaro Fukuda}
\affiliation{Department of Physics, Keio University,
             Kanagawa 223-8522, Japan}
\author{Kenji Fukushima}
\affiliation{Department of Physics, Keio University,
             Kanagawa 223-8522, Japan}
\author{Tomoya Hayata}
\affiliation{Department of Physics, The University of Tokyo,
             Tokyo 113-0031, Japan}
\affiliation{Theoretical Research Division, Nishina Center,
             RIKEN, Saitama 351-0198, Japan}
\author{Yoshimasa Hidaka}
\affiliation{Theoretical Research Division, Nishina Center,
             RIKEN, Saitama 351-0198, Japan}

\begin{abstract}
 We investigate the sign problem of the fermion determinant at finite
 baryon density in (1+1) dimensions, in which the ground state in the
 chiral limit should be free from the sign problem by forming a chiral
 spiral.  To confirm it, we evaluate the fermion determinant in the
 continuum theory at the one-loop level and find that the determinant
 becomes real as expected.  The conventional lattice formulation to
 implement a chemical potential is, however, not compatible with the
 spiral transformation.  We discuss an alternative of the
 finite-density formulation and numerically verify the chiral spiral
 on the finite-density lattice.
\end{abstract}
\maketitle


\paragraph{Introduction}
Quantum chromodynamics (QCD) has profound contents to be explored with
external parameters such as the temperature $T$, the baryon chemical
potential $\muB$ (that is equal to the quark chemical potential $\muq$
multiplied by the number of colors $\Nc$), the magnetic field $B$, and
so on~\cite{review}.  The direct calculation based on QCD would be,
however, feasible only in some limited ranges of these parameters.  In
particular along the direction of increasing $\muq$, perturbative QCD
is not really useful unless the density is high enough to accommodate
color superconductivity~\cite{Rischke:2000cn,Alford:2007xm}.
Moreover, the numerical simulation based on lattice QCD breaks down
with finite $\muq$.  The most serious obstacle lies in the fact that
the Monte-Carlo method based on importance sampling is invalid for the
finite-density case due to the complex fermion determinant, which is
commonly referred to as the sign problem (see \cite{Muroya:2003qs} for
reviews).

The sign problem is relevant not only in the lattice-QCD simulation
but also in analytical
computations~\cite{Dumitru:2005ng,Fukushima:2006uv}.  At finite
temperature, the temporal or thermal component of the gauge field 
$A_4$ plays a special role, and its expectation value is given a
gauge-invariant interpretation, namely, (the phase of) the Polyakov
loop, $\Pol\equiv\calP\exp[ig\int_0^\beta dx_4 A_4]$.  Because the
traced Polyakov loop is an order parameter for quark deconfinement,
many efforts have been devoted to the computation of the effective
potential with respect to $\Pol$ or
$A_4$~\cite{Gross:1980br,Weiss:1980rj,KorthalsAltes:1993ca}.  With the
contribution from the fermion determinant~\cite{Belyaev:1991np}, the
effective potential at nonzero $\muq$ has turned out to take a
complex value, and thus the physical meaning as a grand potential or
thermal weight is obscure.  This is how we can observe the sign
problem even using the perturbative calculation in the continuum
theory.

Since the resolution of the sign problem seems to be still far from
our hands, it is very instructive to acquire some experiences with
density-like effects that would cause no sign problem.  Theoretical
attempts along this line include the imaginary chemical
potential~\cite{ImaginaryChemicalPotential}, the isospin chemical
potential~\cite{Son:2000xc}, the chiral chemical
potential~\cite{Yamamoto:2011gk}, dense QCD with two colors or with
adjoint matter~\cite{Muroya:2003qs,Kogut:2000ek}, the strong magnetic
field $B$~\cite{D'Elia:2010nq}, and so on.  Among these examples the
magnetic field particularly leads to a quite suggestive change in the
state of quark matter.  The most drastic consequences result from the
Landau quantization and the dominance of the lowest Landau level for
spin-$1/2$ fermions.  Thus, in the strong-$B$ limit, quarks are
subject to the dimensional reduction, and the transverse motion in a
plane perpendicular to $B$ is frozen.

The nature of chiral symmetry breaking is affected accordingly by the
strong-$B$ effects~\cite{Suganuma:1990nn};  the spontaneous breaking
of chiral symmetry inevitably occurs in the (1+1)-dimensional system
(or in the lowest Landau level approximation~\cite{Gusynin:1994re}), 
which is called the magnetic catalysis.  This phenomenon is analogous
to the superconductivity, which is also triggered by the
low-dimensional nature on the Fermi surface.  In chiral model studies
(see \cite{Gatto:2012sp} for a review) the chiral phase transition is
delayed toward a higher temperature due to the magnetic catalysis,
while in the finite-$T$ lattice-QCD simulation it has been recognized
that the chiral crossover temperature gets smaller with increasing
$B$, which is sometimes called the inverse magnetic catalysis or the
magnetic inhibition~\cite{Fukushima:2012kc}.  Another interesting
example from the (1+1)-dimensional nature is the topological
phenomenon such as the chiral magnetic effect~\cite{Kharzeev:2007jp}
that might be detectable with the noncentral collision of positively
charged heavy ions through charge separation or photon
emission~\cite{Hattori:2012je}.  The quickest derivation of the chiral
magnetic effect makes use of the low dimensionality of the Landau
zero-mode, that is, the special property of the $\gamma$ matrices;
$\gamma^5\gamma^\mu=\epsilon^{\mu\nu}\gamma^\nu$, in (1+1) dimensions.

Such a (1+1)-dimensional system of quark matter provides us with further
useful information;  the ground state of the (1+1)-dimensional chiral
system at finite density (with large number of internal degrees of
freedom) is known to form a chiral spiral~\cite{ChiralSpiral} (see
also \cite{deForcrand:2006zz} for a lattice study).  In the strong-$B$
limit, therefore, the ``chiral magnetic spiral'' could be one of the
most likely candidates for the ground state of finite-density and
magnetized quark matter~\cite{Basar:2010zd}.  The essential idea of
\cite{Basar:2010zd} is that the explicit $\muq$ dependence is rotated
away in (1+1) dimensions, and this procedure transforms the
homogeneous chiral condensate to form a spiral in chiral basis.  This
at the same time implies that the sign problem should no longer be
harmful once the dimensional reduction occurs.

One might have thought that the strong-$B$ limit is such a special
environment having loose relevance to our realistic world.  It has
been argued, however, that quark matter at high density even without
$B$ already exhibits a character as a pseudo-(1+1)-dimensional system
locally on the Fermi surface~\cite{Kojo:2009ha}, just like the
situation of superconductivity, and the whole Fermi surface should be
covered by low-dimensional patches~\cite{Kojo:2011cn}.  Besides, the
$p$-wave pion condensation in nuclear matter having the same spiral
structure is still a vital possibility beyond the normal nuclear
density~\cite{Tatsumi:2003fa}.  In this way, it is definitely worth
considering the sign problem and the ground state structure in
(1+1)-dimensional systems both for academic interest and for practical
purpose.

Our analysis in the present work surprisingly reports that the
conventional lattice formulation at finite density becomes problematic
even for describing the expected ground state of such an idealized
(1+1)-dimensional system.  First we shall illuminate how the sign
problem should be irrelevant in (1+1) dimensions by performing the
perturbative calculation.  Then, we will proceed to the lattice
formulation to find that the conventional introduction of
$\muq$~\cite{Hasenfratz:1983ba} cannot realize the transformation
properties in the continuum theory unless the lattice spacing is very
small.  We can choose an alternative that is optimal to yield a chiral
spiral and conduct the numerical test to confirm a spiral formation on
the lattice.
\vspace{0.3em}


\paragraph{Perturbative calculation}
Let us first evaluate the fermion determinant at finite $\muq$ and
high enough $T$ that justifies the perturbative treatment.  At the
one-loop level in the deconfined phase, we should keep the
Polyakov-loop $A_4$ background and carry out the Gaussian integration
with respect to quantum fluctuations of gluons.  After taking the
summation over the Matsubara frequency, we can write the determinant
$\calM[A_4]$ (for a single flavor throughout this work) down as
\begin{equation}
 \begin{split}
 & \calM[A_4] = \calN \rme^{-T^d V_d \Gamma[A_4]}
  = \calN \exp\Biggl\{\alpha_d \int\frac{V_d d^d p}{(2\pi)^d} \\
 &\qquad  \tr\ln\Bigl[\bigl( 1\!+\! \Pol
  \rme^{-(\varepsilon-\muq)/T} \bigr) \bigl( 1\!+\! \Pol^\dagger
  \rme^{-(\varepsilon+\muq)/T} \bigr) \Bigr] \Biggr\} \;,
 \end{split}
\label{eq:det}
\end{equation}
where $d$ and $V_d$ represent the spatial dimension and the spatial
volume, respectively, and the dispersion relation is
$\varepsilon = \sqrt{p^2 + m^2}$.  We note that the spin degeneracy
factor $\alpha_d$ depends on $d$:  $\alpha_3=2$ and $\alpha_1=1$.  For
practical convenience, we rotate the color basis as
$U\Pol U^\dagger=\text{diag}(\rme^{\rmi\pi\phi_1},\rme^{\rmi\pi\phi_2},
\rme^{\rmi\pi\phi_3})$, where $\phi_1+\phi_2+\phi_3=0$ should hold to
satisfy $\det(U\Pol U^\dagger)=1$.

For the massless case ($m=0$), we can perform the full analytical
integration for arbitrary $d$.  In particular, a choice of $d=3$
immediately yields the well-known Weiss--Gross-Pisarski-Yaffe-type
potential~\cite{Weiss:1980rj} that takes the following polynomial
form~\cite{review,Belyaev:1991np,Roberge:1986mm}:
\begin{equation}
 \Gamma[\phi] = -\frac{4\pi^2}{3} \sum_{i=1}^{\Nc}
  B_4\biggl[ \biggl(\frac{1+\phi_i}{2}\biggr)_{\!\!\mathrm{mod 1}}
  -\frac{\rmi\tmu}{2} \biggr] \;,
\label{eq:weiss}
\end{equation}
where the Bernoulli polynomial appears as
$B_4(x)=x^2(1-x)^2-1/30$~\cite{KorthalsAltes:1993ca}.  We also
introduced the dimensionless chemical potential as
$\tmu\equiv \muq/(\pi T)$ for notational simplicity.  While we choose
$\Nc=3$ in our QCD study, Eq.~\eqref{eq:weiss} is valid for any
$\mathrm{SU}(\Nc)$ groups.

\begin{figure}
 \includegraphics[width=0.8\columnwidth]{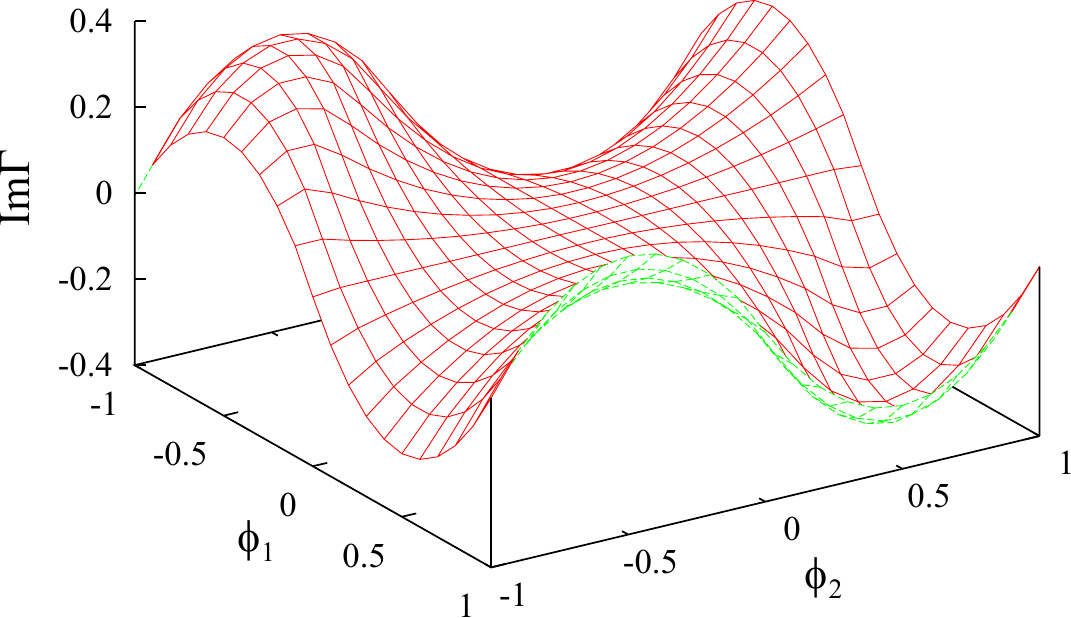} \\
 \includegraphics[width=0.8\columnwidth]{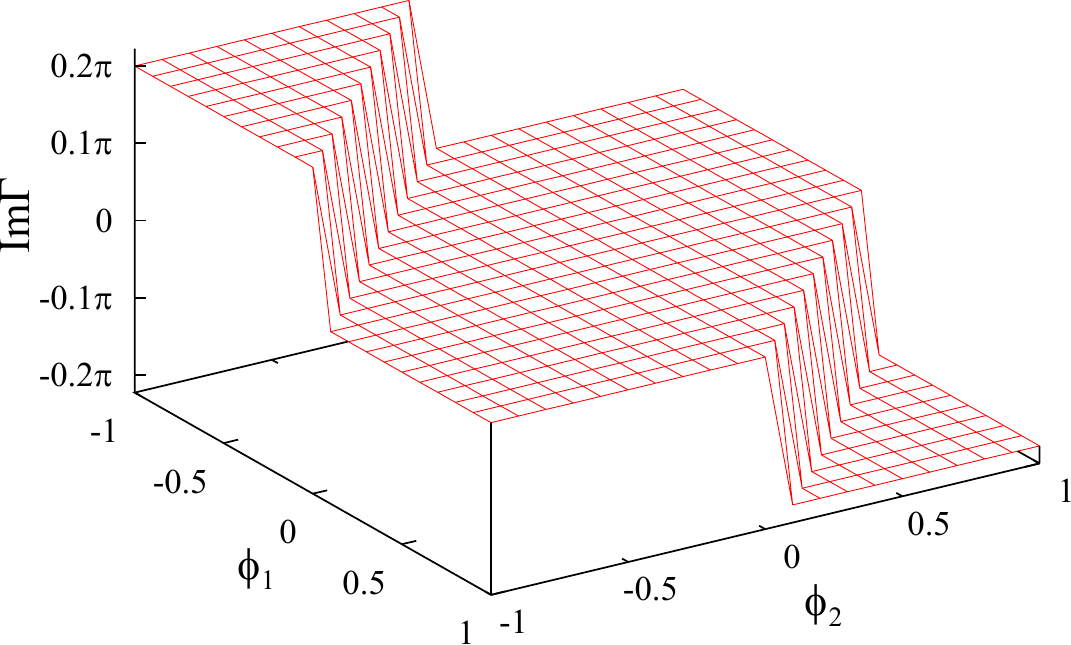}
 \caption{(Upper) Imaginary part of $\Gamma[\phi]$ for $d=3$ at
   $\tmu=0.1$ shown as a function of $\phi_1$ and $\phi_2$.\ \ (Lower)
   Imaginary part of $\Gamma[\phi]$ for $d=1$ at $\tmu=0.1$.}
 \label{fig:poten}
\end{figure}

The apparent presence of the imaginary part in Eq.~\eqref{eq:weiss}
corresponds to the sign problem.  Indeed, the complex phase of the
fermion determinant is nothing but $-T^d V_d\text{Im}\Gamma$
(mod $2\pi$).  To gain a more informative view, we make a plot for
$\text{Im}\Gamma[\phi]$ in the upper panel of Fig.~\ref{fig:poten} as
a function of $\phi_1$ and $\phi_2$ (with $\phi_3=-\phi_1-\phi_2$).
It is clear from the figure that the complex phase has a nontrivial
dependence on the gauge configuration $\phi_1$ and $\phi_2$.

A more interesting case is for $d=1$ corresponding to quark matter
under the dimensional reduction.  In this case the logarithm of the
fermion determinant simplifies as
\begin{equation}
 \Gamma[\phi] = 2\pi \sum_{i=1}^{\Nc} B_2 \biggl[
  \biggl(\frac{1+\phi_i}{2}\biggr)_{\!\!\mathrm{mod 1}}
  -\frac{\rmi\tmu}{2} \biggr] \;.
\label{eq:weiss2}
\end{equation}
Here, we again used the Bernoulli polynomial as defined by
$B_2(x)=x^2-x+1/6$.  It is quite reasonable that $B_4(x)$ in
Eq.~\eqref{eq:weiss} for $d+1=4$ is replaced with $B_2(x)$ in
Eq.~\eqref{eq:weiss2} for $d+1=2$.  We can identify the imaginary part
for $d=1$ as
\begin{equation}
 \begin{split}
 \text{Im}\Gamma[\phi] &=  2\pi \tmu\sum_{i=1}^{\Nc} \Biggl[
  \biggl(\frac{1+\phi_i}{2}\biggr)_{\mathrm{mod 1}} -
  \frac{1}{2}\biggr] \\
 &= \left\{ \begin{array}{cp{1em}l}
  0 && (-1 < \phi_1 + \phi_2 < 1) \\
  -2\pi\tmu  && (\phi_1 + \phi_2 \ge 1) \\
  +2\pi\tmu && (\phi_1 + \phi_2 \le -1)
  \end{array} \right. \;.
 \end{split}
\label{eq:imG}
\end{equation}
This analytic behavior is visually shown in the lower panel of
Fig.~\ref{fig:poten}.  The step emerges when
$(1+\phi_3)/2=(1-\phi_1-\phi_2)/2$ exceeds the boundary of modular
one, and then, as is clear from the above expression,
$\text{Im}\Gamma$ takes a constant $\pm2\pi\tmu$, which is
$\pm0.2\pi$ in our numerical setup ($\tmu=0.1$) to draw
Fig.~\ref{fig:poten}.

With a more careful deliberation on the phase-space volume, we see
that an imaginary part in the region $|\phi_1+\phi_2|\ge 1$ has no
contribution since this finite value is quantized as
$T V_1\text{Im}\Gamma = 2\pi n$ with an integer $n$.  To see this, let
us consider the branch-cut contribution from the logarithm in the
integrand of Eq.~\eqref{eq:det}, which appears when the real part in
the logarithm turns negative, i.e.,
$\mathrm{Re}[\rme^{\rmi\pi\phi_i-(|p|-\muq)/T}]<-1$.  The momentum
integration under this condition picks up the phase-space volume
satisfying $|p|<\muq$, that is, 
\begin{equation}
 \begin{split}
 & \pm \int \frac{V_1 dp}{2\pi} 2\pi\,\theta(\muq-|p|) \\
 \to & \pm \!\!\!\!\!\!\!\! \sum_{\text{Phase Space}} \!\!\!\!\!\!\!
  2\pi\,\theta(\muq-|p|) = \pm 2\cdot 2\pi \biggl\lfloor
  \frac{V_1\muq}{2\pi} \biggr\rfloor \;,
 \end{split}
\label{eq:phasespace}
\end{equation}
where $\lfloor\cdots\rfloor$ represents the floor function.
Equation~(\ref{eq:phasespace}) reproduces Eq.~\eqref{eq:imG}
multiplied with $T V_1$ in a quantized form.  In summary of
perturbative analyses in the (1+1)-dimensional continuum theory, as
conjectured, we have actually confirmed that no sign problem arises.
\vspace{0.3em}


\paragraph{Lattice formulation}
This simple analytical observation is, however, not easy to be
validated on the lattice, unless one reaches the continuum limit.  To
make the point explicitly clear, let us take a
pseudo-(1+1)-dimensional system discarding two transverse (1st and
2nd) components.  Then, in Euclidean space-time with the longitudinal
(3rd) and the temporal (4th) components, the Lagrangian density,
$\calL_{\text{eff}}=\bar{\psi} D(\muq) \psi$, with $D(\muq) =
\gamma^3(\partial_3-\rmi g A_3)+\gamma^4(\partial_4+\muq-\rmi g A_4)$
defines the theory.  Here, we consider the most interesting case of
$m=0$ only.  Then, we can immediately confirm that $\muq$ is
superficially erased by the following rotation:
\begin{equation}
 \psi \;\to\; \psi = U\psi' \;,\qquad
 \bar{\psi} \;\to\; \bar{\psi} = \bar{\psi}' U \;,
\label{eq:trans}
\end{equation}
with $U\equiv\exp\bigl( -\muq\gamma^3\gamma^4 x_3 \bigr)$.  The
chemical potential can be factorized out by the unitary
transformation, $D(\muq)=U^\dag D(0) U^\dag$, and thus the fermion
determinant is independent of $\muq$.  Strictly speaking, this
rotation also causes a shift in momenta carried by $\psi'$ and
$\bar{\psi}'$, and such a shift gives rise to nontrivial $\muq$
dependence through chiral anomaly~\cite{ChiralSpiral}.  For the
moment, it suffices for our purpose of seeing the spiral if we focus
on the tree-level elimination of the $\muq$-term, and we will not go
into anomalous $\muq$ dependence.

In the conventional lattice formulation~\cite{Hasenfratz:1983ba},
$\muq$ is introduced as
\begin{equation}
 \begin{split}
 & \bar{\psi}\gamma^4(\partial_4 + \muq)\psi
  = (\bar{\psi}\,\rme^{-\muq x_4}) \gamma^4\partial_4 (\rme^{\muq x_4}\psi)
  \\
 & \simeq \frac{1}{2}\bar{\psi}(x) \gamma^4 \rme^{\muq}\psi(x+\hat{4})
  -\frac{1}{2}\bar{\psi}(x) \gamma^4 \rme^{-\muq}\psi(x-\hat{4}) \;.
 \end{split}
\label{eq:incomplete}
\end{equation}
If we apply the transformation with Eq.~\eqref{eq:trans} on the
lattice version of the Lagrangian, we can find $U D(\muq) U$ as
\begin{align}
 &\frac{1}{2}\biggl\{  \bar{\psi}'(x) \Bigl[ (\gamma^3 \cos\muq
  \!-\!\gamma^4\sin\muq)\psi'(x\!+\!\hat{3})
  \!+\!\gamma^4  \rme^{\muq}\psi'(x\!+\!\hat{4}) \Bigr] \notag\\
 & - \bar{\psi}'(x) \Bigl[(\gamma^3 \cos\muq \!+\! \gamma^4 \sin\muq)
  \psi'(x\!-\!\hat{3}) \!+\! \gamma^4 \rme^{-\muq}\psi'(x\!-\!\hat{4})
  \Bigr] \biggr\}
\end{align}
apart from the link variables.  In the continuum limit (i.e.\ the
lattice spacing $a\to0$), where $\muq a$ goes vanishingly small, the
explicit $\muq$ dependence certainly disappears as anticipated from
the continuum theory.  In this sense, such an incomplete cancellation
in Eq.~\eqref{eq:incomplete} is a lattice artifact, and yet, this
is crucial for the sign problem and the formation of chiral spiral.

One quick remedy for the noncancellation problem is to alter the way
to formulate $\muq$ on the lattice.  We shall propose to introduce the
chemical potential as $D(\muq)=U^{\dag}D(0)U^{\dag}$, i.e.\ (see
\cite{Creutz:2010cz} for a similar proposal),
\begin{align}
 &\bar{\psi}\bigl(\gamma^3\partial_3+\gamma^4\muq\bigr)\psi
  = (\bar{\psi}\,U^\dag)\gamma^3\partial_3
  (U^\dag\psi) \notag\\
 &\simeq \frac{1}{2}\bar{\psi}(x) \bigl(\gamma^3 \cos\muq
  +\gamma^4 \sin\muq\bigr) \psi(x+\hat{3}) \notag\\
 &\quad -\frac{1}{2}\bar{\psi}(x) \bigl(\gamma^3 \cos\muq
  -\gamma^4 \sin\muq\bigr) \psi(x-\hat{3})
\label{eq:new}
\end{align}
(with link variables omitted).  In this form, it may look trivial at
glance that the rotation with Eq.~\eqref{eq:trans} can get rid of the
$\muq$ dependence.  The situation is not such trivial, though.  One
can actually prove that the eigenvalues of this fermion operator
appear as a quartet: $\lambda$, $-\lambda$, $\lambda^\ast$, and
$-\lambda^\ast$.  In other words, the fermion determinant is always
real regardless of the dimensionality!  Needless to say, this cannot
be a resolution of the sign problem.  Because $\sin\muq$ is
accompanied by $\cos q_3$ in momentum space of Eq.~\eqref{eq:new}, the
sign of $\muq$ changes for the fermion doublers in the $\hat{3}$
direction.  Therefore, if we interpret the doublers as different quark
flavors, $\muq$ as put in Eq.~\eqref{eq:new} represents the isospin
chemical potential rather than the quark chemical potential, so that
the determinant is always real!  This also means that the new
formulation as in Eq.~\eqref{eq:new} cannot produce a chiral spiral.

Thus, we must cope with the doubler problem to treat $\muq$ as a 
quark chemical potential.  In this work we shall na\"{i}vely add the
Wilson term, $r_W \bar{\psi}\partial^2 \psi$ (where we choose
$r_W=1$) to make heavy doublers decouple from the dynamics.  Not to
violate the transformation properties, $D(\muq)=U^\dag D(0)U^\dag$, we
must implement the Wilson term according to Eq.~\eqref{eq:new} as
$r_W\bar{\psi}\partial^2\psi \rightarrow r_W
(\bar{\psi}U^\dag)\partial^2 (U^\dag \psi)$.  In this case the fermion
determinant becomes real only for discrete values of $\muq$, that is
quantized as $\muq=(\pi/N)n$, where $N$ is the number of lattice sites
along the $x_3$ direction.  Because the Wilson term has an explicit
$x_3$ dependence, we must require $\rme^{2\muq\gamma^3\gamma^4 N}=1$
to keep the action invariant under the shift: $x_3\to x_3+N$.

\begin{figure}
 \includegraphics[width=0.77\columnwidth]{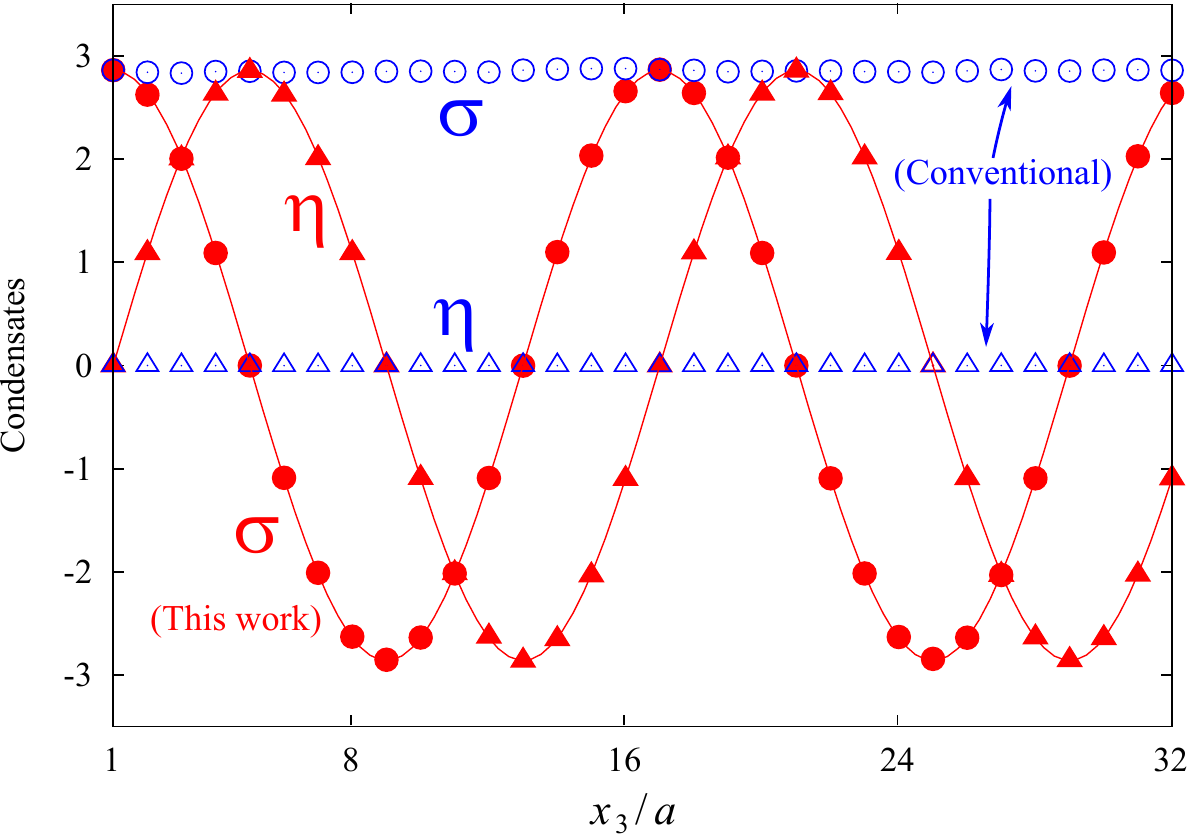}
 \caption{Condensates $\sigma$ and $\eta$ at $\muq=2\pi/N$ (with
   $N=32$) as a function of $x_3$ in the lattice unit.  The closed
   circle and triangle dots represent the results from our new
   formulation, while the open circle and triangle dots from the
   conventional one.  The solid curves are $2.86\cos[2\muq (x_3-1)]$
   and $2.86\sin[2\muq (x_3-1)]$ that fit the oscillation behavior.}
 \label{fig:spiral}
\end{figure}

For $\muq=(\pi/N)n$, the determinant returns to a real value.  While
the bulk properties are fixed by the whole quantity of the
determinant, we emphasize here, the microscopic dynamics is far more
nontrivial.  If the vacuum at $\muq=0$ has a nonzero and homogeneous
chiral condensate $\sigma_0\equiv\langle\bar{\psi}\psi\rangle\neq0$,
the rotated vacuum with Eq.~\eqref{eq:trans} at $\muq\neq0$ should
yield $\sigma_0 = \langle\bar{\psi}'\psi'\rangle$ as well.  In terms
of the original basis, accordingly, we can expect
$\sigma\equiv\langle\bar{\psi}\psi\rangle=\sigma_0 \cos(2\muq x_3)$
and
$\eta\equiv\langle\bar{\psi}\gamma^3\gamma^4\psi\rangle=\sigma_0
\sin(2\muq x_3)$, which locally breaks chiral symmetry but does not
globally, i.e., the average of the condensate vanishes:
$\int d^2x\langle\bar{\psi}\psi\rangle=0$.

In Fig.~\ref{fig:spiral} we show the condensates as a function of
$x_3$ defined by
$\sigma(x_3)\equiv N_t^{-1}\sum_{x_4} \text{tr}[D^{-1}(\muq)]$ and
$\eta(x_3)\equiv N_t^{-1}\sum_{x_4} \text{tr}[\gamma^3\gamma^4
D^{-1}(\muq)]$ (in the lattice unit).  This is the result for one
gauge configuration generated after 1000 quench updates using the
Wilson gauge action with $\beta=5.0$.  If we use the conventional
introduction of $\muq$ as in Eq.~\eqref{eq:incomplete}, only $\sigma$
has a finite expectation value and the oscillatory pattern is hardly
visible.  With the new formulation as in Eq.~\eqref{eq:new}, on the
other hand, both $\sigma$ and $\eta$ take a finite value to develop a
clear chiral spiral.  [One should be careful to interpret this
  result:  The exact chiral limit with strict (1+1) dimensions gives
  rise to no chiral condensate.  This is why we set our problem in
  pseudo-(1+1) dimensions and also the Wilson term plays a role.]

\begin{figure}
 \includegraphics[width=0.65\columnwidth]{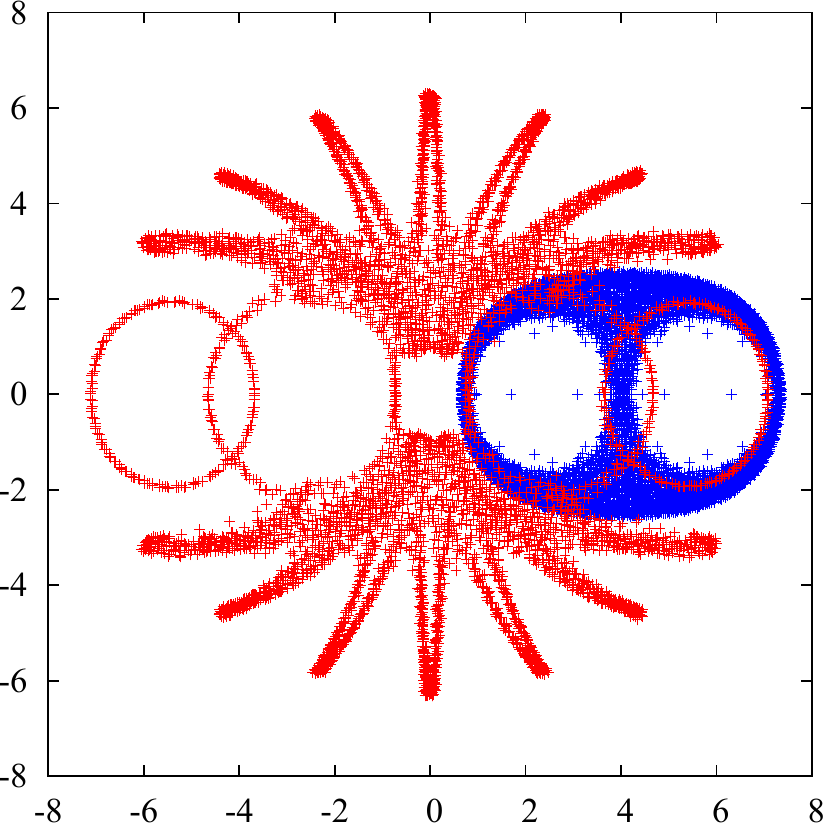}
 \caption{Eigenvalue distribution of the finite-$\muq$ fermion
   operator on the $32\times 32$ lattice ($N=32$) for the gauge
   configuration corresponding to Fig.~\ref{fig:spiral}.  Results with
   $\muq=2\pi/N$ (red dots) overlaid on those with $\muq=0$ (blue
   dots).}
 \label{fig:ev}
\end{figure}

Since the chiral condensate is related to the low-lying eigenvalues
via the Banks-Casher relation, it is interesting to see how the
eigenvalue distribution changes with the chiral spiral.  The Wilson
term breaks anti-Hermiticity, and the eigenvalues are complex even at
$\muq=0$, so that the original Banks-Casher relation needs a
modification;  the chiral condensate should be derived from the
eigenvalues of $D^\dag(0)D(0)$ rather than
$D(0)$~\cite{Giusti:2008vb}.  In this work, we do not calculate the
former, and yet, it is quite interesting to investigate the
qualitative changes of the latter at finite $\muq$, which is presented
in Fig.~\ref{fig:ev}.

Figure~\ref{fig:ev} shows the eigenvalue distribution of $D(\muq)$ for
$\muq=0$ (blue dots) and $\muq=2\pi/N$ (red dots) as introduced in
Eq.~\eqref{eq:new}.  At $\muq=0$ the eigenvalue distribution is just
the same as a conventional one.  With increasing $\muq$ the
distribution spreads to the negative real region, and when $\muq$
reaches a multiple of $\pi/N$, the determinant should be identical to
the $\muq=0$ value, though the eigenvalue distribution looks totally
different.  Although the distribution appears to be symmetric for
$\muq=(2\pi/N)n$ as seen in Fig.~\ref{fig:ev}, there is no longer a
quartet structure nor any pairwise symmetry.  It is miraculous that
the product of all these eigenvalues happens to be real.
\vspace{0.3em}


\paragraph{Conclusions}
We have justified the idea that the sign problem of the fermion
determinant at finite $\muq$ be irrelevant in the (1+1)-dimensional
system.  This is caused by the chiral transformation that removes the
chemical potential.  We have first evaluated the determinant
perturbatively in the continuum theory, and found that the imaginary
part in the (1+1)-dimensional case vanishes unlike the
(3+1)-dimensional situation that suffers from the sign problem.

For the discretized fermion on the (1+1)-dimensional lattice, the
conventional way to impose a chemical potential causes the sign
problem, which is a lattice artifact and should be absent in the
continuum limit.  In practice, however, this lattice artifact severely
hinders the formation of the chiral spiral.  To evade this problem, we
have proposed a new method to introduce a chemical potential by
twisting the Dirac operator along one of the spatial directions by
$\exp(\muq\gamma_3\gamma_4 x_3)$, which recovers the correct
continuum limit as it should.  In this case, the fermion determinant
becomes real, but it turns out that such a chemical potential induces
not a quark density but a doubler (or isospin) density if the doublers
are not killed.  We then find no chiral spiral.  By diminishing
spurious symmetry with doublers, we have successfully confirmed a
clear chiral spiral.  The eigenvalues of our fermion operator has a
peculiar distribution structure, which suggests some relation between
the appearance of some distribution pattern and the formation of the
chiral spiral, which we leave for a future problem.

Our idea of the spatially twisted chemical potential can be applied to
not only (1+1) dimensions but also more general dimensions.  If the
spiral structure is the genuine ground state at strong magnetic field
or at high baryon density that brings about the dimensional reduction,
the conventional formulation with $\muq$ is not really an optimal
choice.  The present work has manifestly demonstrated the advantage of
the new formulation to investigate the sign problem and the chiral
spiral.  It is also an intriguing future problem to study our method
using the other fermions, particularly the overlap fermion that also
exhibits a peculiar distribution of finite-density
eigenvalues~\cite{Bloch:2006cd}.

\acknowledgments
We thank A.~Yamamoto for stimulating discussions and helpful
comments.  We acknowledge the Lattice Tool Kit (LTKf90) with which we
generated the gauge configuration.  T.~H. was supported by JSPS
Research Fellowships for Young Scientists.  This work was supported by
RIKEN iTHES Project, and JSPS KAKENHI Grants Numbers 24740169,
24740184, and  23340067.


\end{document}